\documentclass{article}
\usepackage{spadre2008}
\usepackage{graphicx}
\usepackage{amssymb}
\frompage{000} \topage{000}                                              %|
\def\rightmark{The Electron-Ion Collider}
\def\leftmark{C. A. Aidala for the EIC Collaboration}

\title{Peering Into Hadronic Matter: The Electron-Ion Collider}
\authors{
{Christine A. Aidala$^{1,a}$ for the EIC Collaboration$^2$ %
}\\[2.812mm]
{\normalsize
\hspace*{-8pt}$^1$ University of Massachusetts, \\
Amherst, MA 01003, U.S.A.\\[0.2ex]
\hspace*{-8pt}$^2$ http://web.mit.edu/eicc/
}}

\abstract{A unique new facility, capable of colliding beams of electrons with a wide range of nuclei as well as
polarized protons and light ions, has been proposed to study the role of
gluons in matter and perform precision mapping of the structure of nucleons
and nuclei.  The physics prospects of the proposed Electron-Ion Collider (EIC) are discussed.}

\keyword{Quantum chromodynamics, DIS structure functions, nucleon spin, DIS off nuclei, saturation, EIC}
\PACS{13.60.Hb, 24.85.+p, 14.20.Dh, 13.87.Fh}

\begin{document}

\maketitle
\setcounter{page}{1}

\section{Introduction}\label{intro}

The intellectual drive of mankind to understand the nature of everyday matter is one that goes back to ancient times.  The discovery of the atom in the nineteenth century and subsequently the basic building blocks of atoms--protons, neutrons, and electrons--during the first half of the twentieth century represented great leaps forward.  The partonic substructure of protons and neutrons became evident in the 1960's, and the 1970's saw the development of quantum chromodynamics (QCD), the theory of the strong force governing the interactions among the partons within hadronic matter.  Within QCD colored, self-interacting gluons serve as the mediators of the strong force, with gluon interactions responsible for approximately 98\% of the mass of the visible universe.  However, as elegant and powerful a theory as QCD may be, there is much it does not reveal.  It does not tell us in any detail about the partonic landscape within hadronic matter, nor about the process by which quarks and gluons fragment into final-state, colorless hadrons.  To investigate the structure of protons and nuclei--the everyday matter of the world around us--in terms of their partonic degrees of freedom, we must presently turn to experiment.

An experimental tool-of-the-trade for the past four decades in investigating the structure of protons and nuclei has been deep-inelastic scattering (DIS) of leptonic probes off of hadronic targets.  The total invariant cross section for deep-inelastic scattering of electrons off of protons is given by:

\begin{eqnarray*}
\frac{d^2 \sigma^{ep \rightarrow eX}}{dx dQ^2} =
\frac{4 \pi \alpha^2_{e.m.}}{xQ^4} \left[ \left(1-y+\frac{y^2}{2} \right )
    F_2(x,Q^2) - \frac{y^2}{2} F_L(x,Q^2) \right]
\end{eqnarray*}
where $x$ is the fraction of the proton momentum carried by the struck parton, $Q$ is the four-momentum transfer in the scattering, and $y$ is the fractional energy lost by the electron in the proton rest frame.  $F_2$ and $F_L$ represent inclusive structure functions.  The former is sensitive to the sum of quark and anti-quark momentum distributions; at small $x$, these are the sea quarks. The latter is sensitive to the gluon momentum distribution, which can be accessed via measurements of $F_L$ at different values of $y$ for constant values of $x$ and $Q^2$, possible if one varies the experimental center-of-mass energy.  Given that interactions among the quarks are due to gluons, the gluon momentum distribution can also be determined in inclusive DIS from measurements of $F_2$ via scaling violations, $\frac{\partial F_2}{\partial log Q^2} \propto xg(x,Q^2)$.
In semi-inclusive DIS, one can access the gluon distribution through probes such as open charm or dijet production.  Deep-inelastic scattering is an ideal tool for probing the structure of hadronic matter in that measurement of the scattered electron precisely pins down the values of $x$ and $Q^2$ in the scattering.

Since the 1960's there has been a long series of fixed-target experiments investigating the structure of the proton.  The turn-on of the HERA $e+p$ collider in the early 1990's greatly extended the kinematic coverage for measurements of the linear momentum structure of the proton.  To-date, DIS experiments studying the structure of nuclei as well as the angular momentum (spin) structure of the proton have all utilized fixed targets.  A DIS \emph{collider} facility with capabilities for high-energy \emph{nuclear} as well as \emph{polarized} beams would offer outstanding opportunities to further explore the internal structure of protons and nuclei \cite{AnnRev,eicc}.

\section{Studying Nuclear Structure at Low $x$: The Role of Gluons}

\begin{figure}[tb]
     \begin{center}
     \includegraphics[width=0.42\textwidth]{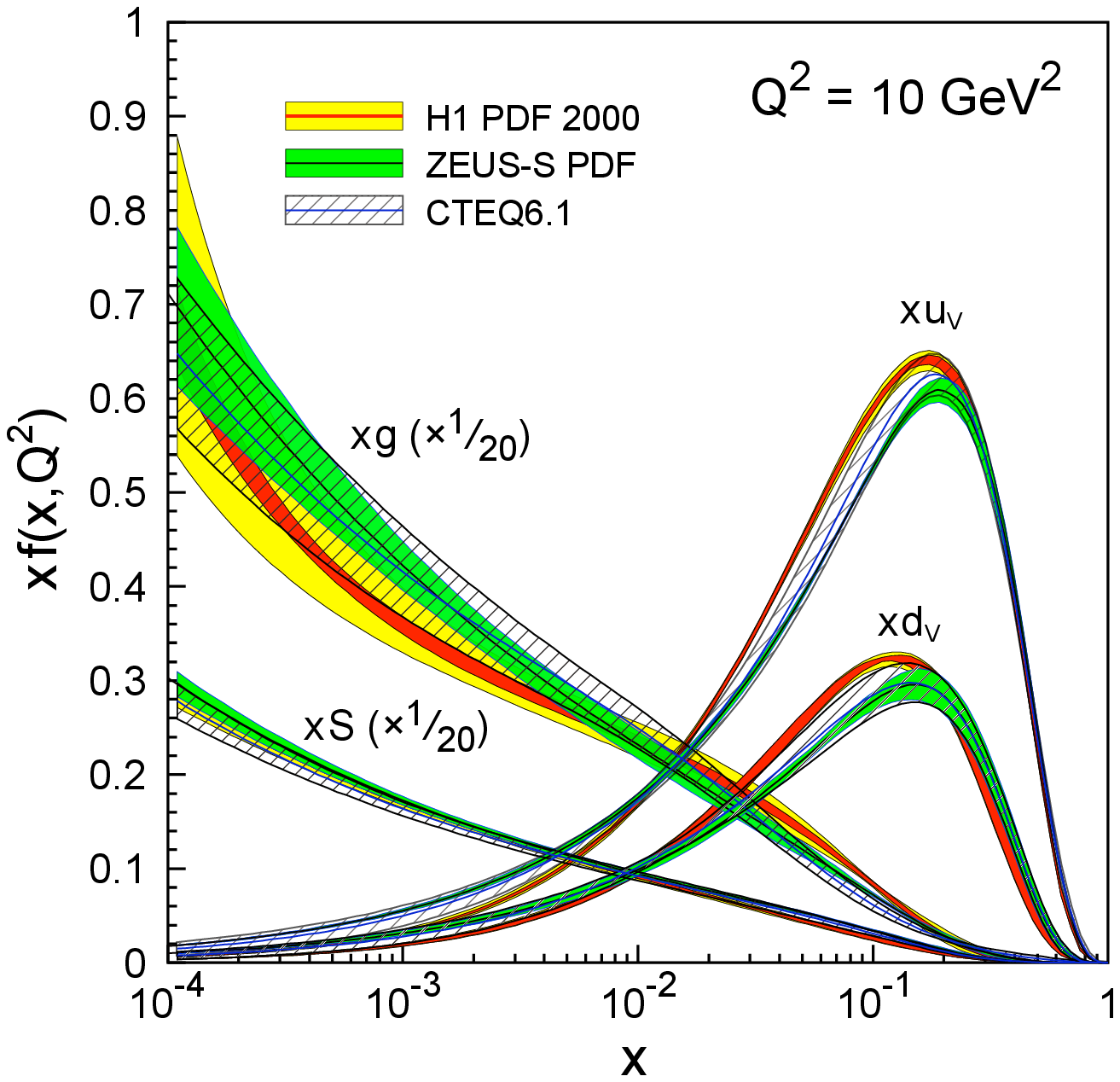}
     \hspace{0.03\textwidth}
     \includegraphics[width=0.48\textwidth]{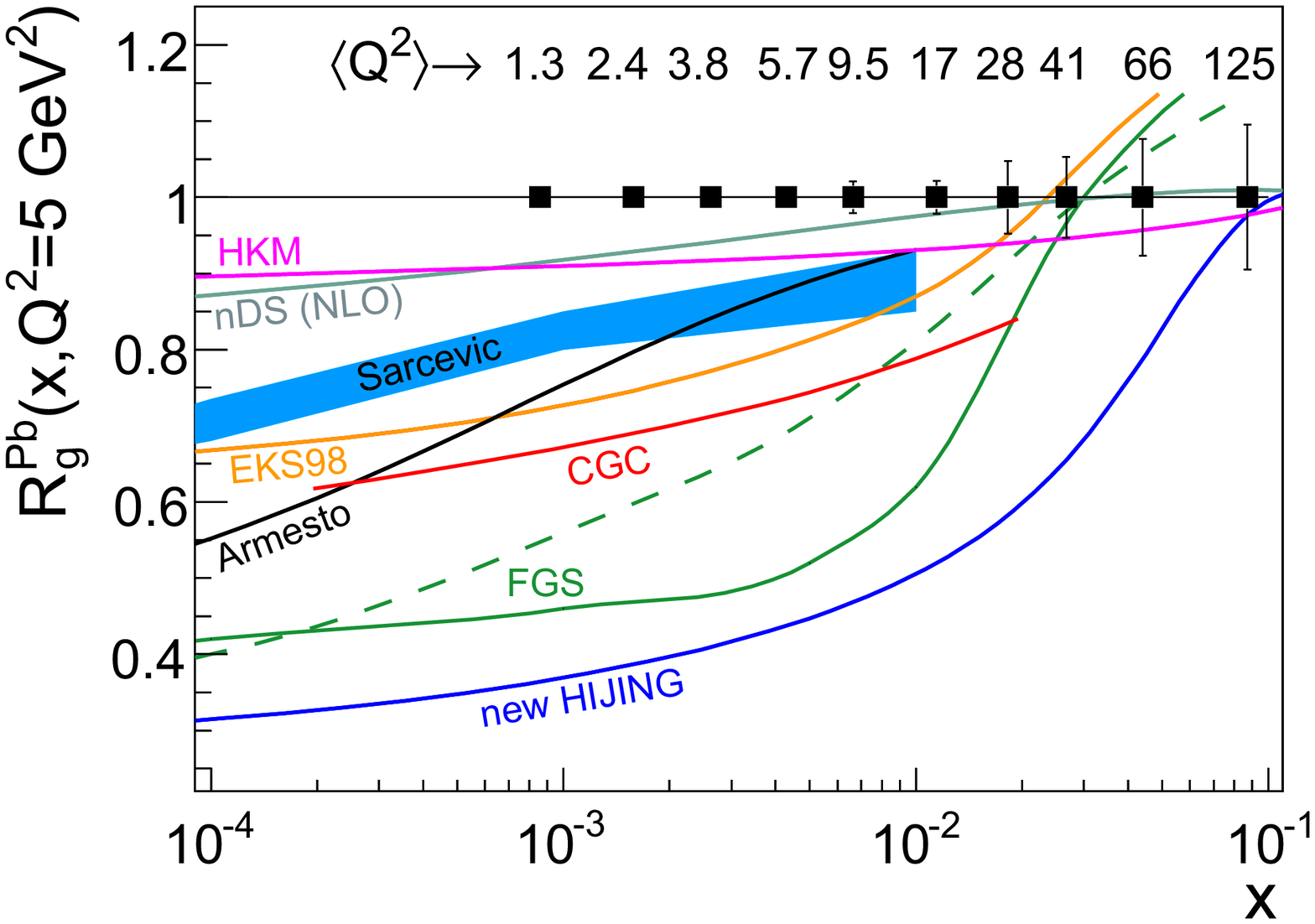}
     \vspace{-7mm}
     \end{center}
     \caption{\textbf{Left:} Gluon, valence, and sea quark momentum
         distributions in the proton obtained from a NLO DGLAP fit to
         $F_2$ measured at HERA \cite{Saxon:2007zz}. Note that the gluon and sea quark
         distributions are scaled down by a factor of 20.
         \textbf{Right:} Predictions for the ratio of the gluon distribution in lead to that in the deuteron and projected EIC statistical uncertainties for 10/A fb$^{-1}$.  }
    \label{fig:pdf}
\end{figure}

Data from the HERA collider have mapped out the linear momentum structure of the proton down to Bjorken $x \approx 6 \times 10^{-5}$.  As can be seen in Fig.~\ref{fig:pdf}, for $Q^2 \gg \Lambda_{\rm QCD}^2$, the gluon density rises quickly for decreasing $x$, and gluons clearly dominate for $x \lesssim  0.01$.  However, despite their abundance at low $x$ and the essential role they play in determining the properties of hadrons, a tremendous amount remains to be learned about gluons in matter.  Within the limited kinematic range covered by previous \emph{nuclear} DIS experiments, striking differences in nuclei with respect to the nucleon have been observed.  Among such nuclear effects is shadowing, a suppression of the nuclear structure function $F^A_2(x,Q^2)$ compared to that of the nucleon, for $x$ values below $\sim 0.1$.  The gluon density in nuclei for $x<0.01$ remains completely unmeasured.  Figure~\ref{fig:pdf} shows predictions for the ratio of the gluon distribution in lead nuclei compared to that in the deuteron, along with projected EIC uncertainties for 10/A fb$^{-1}$.  Given the wide variation in the predictions and small projected experimental uncertainties, measurements at the EIC will provide great insight into the behavior of low-$x$ gluons in nuclei.

In a perturbative QCD (pQCD) framework, experimental measurements of proton or nuclear structure functions at particular values of $Q^2$ and $x$ can be evolved to different values of $Q^2$ and $x$ using the well-established linear evolution equations of DGLAP \cite{Gribov:1972ri,Altarelli:1977zs,Dokshitzer:1977sg} and BFKL \cite{Kuraev:1977fs,Balitsky:1978ic}, respectively.  These sets of evolution equations are applicable for moderate-to-high values of $x$ and $Q^2$; the rapid rise of the gluon density given by these equations as $x$ decreases, which can be understood to be due to the shedding of increasing numbers of low-$x$ gluons by higher-$x$ partons, leads to violation of the "black disc" limit set by the unitarity of the theory.  Gluon saturation mechanisms, allowing small-$x$ gluons to effectively recombine and create partons at higher $x$ values, represent a natural way to control the rising gluon density at low $x$.  The non-linear, small-$x$ renormalization group equations, JIMWLK
\cite{JalilianMarian:1997gr,JalilianMarian:1997dw,Iancu:2000hn,Ferreiro:2001qy}, and the corresponding mean-field BK equations
\cite{Balitsky:1995ub,Kovchegov:1999ua}, can be used to evolve these non-linear effects across a range of energies.

\begin{figure}[tb]
     \begin{center}
     \includegraphics[width=0.5\textwidth]{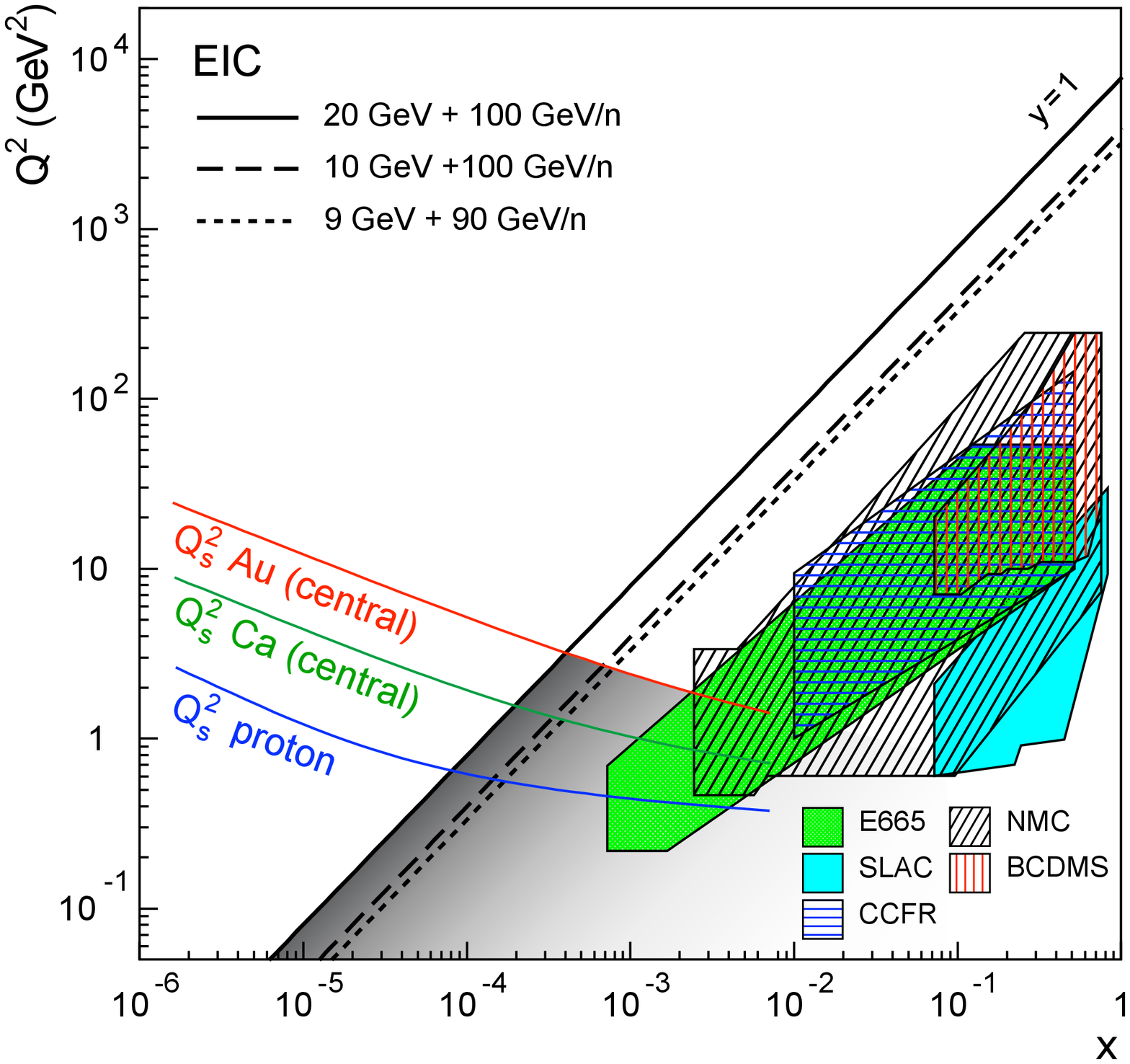}
    \hspace{0.02\textwidth}
    \includegraphics[width=0.46\textwidth] {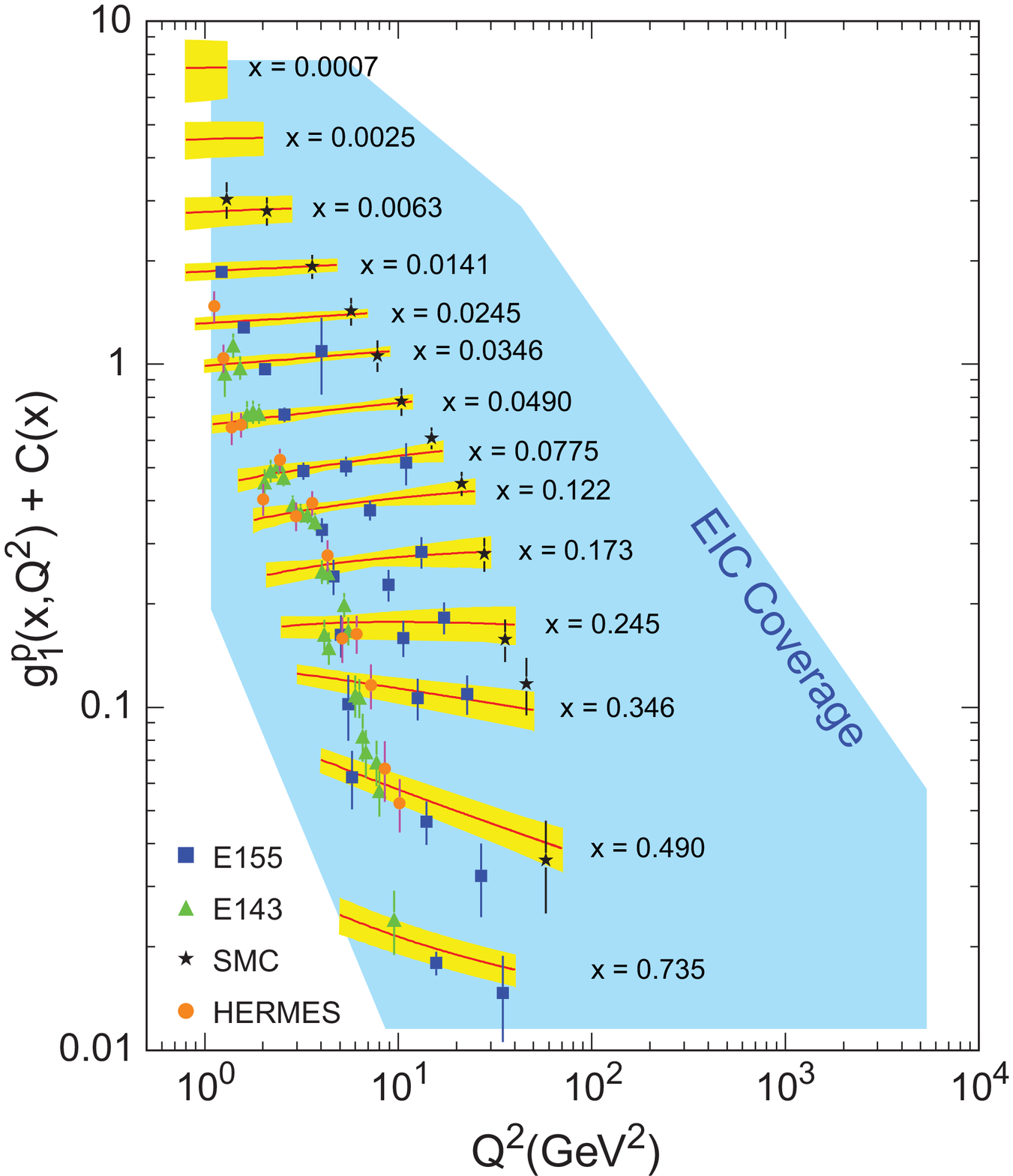}
     \vspace{-10mm}
     \end{center}
     \caption{
         \textbf{Left:} Kinematic acceptance in the ($Q^2,x$) plane for the EIC, shown for various design concepts.  Lines showing the quark saturation scale $Q_s^2$ for protons, Ca, and Au nuclei are superimposed, with the shaded region indicating the range where saturation effects are expected.  The kinematic
         coverage of previous $e$+A, $\mu$+A, and $\nu$+A experiments is indicated.
         \textbf{Right:} Existing measurements of the polarized structure function of the proton, $g_1$, compared to the expected EIC kinematic coverage.
         }
    \label{fig:kinematics}
\end{figure}

The onset of saturation is characterized by a dynamical scale, $Q_s^2$, which depends directly on both energy and nuclear size A.  The enhancement of the saturation scale with increasing A means that the saturation regime can be reached at significantly lower energies in heavy nuclei compared to the nucleon (see Fig.~\ref{fig:kinematics}).  It is also important to note that once in the saturation regime, for $Q^2 < Q_s^2$, the strong coupling $\alpha_s = \alpha_s(Q_s^2)$, i.e.\ for the enhanced values of $Q_s^2$ in a heavy nucleus, perturbative calculations can be performed even at very low values of $Q^2$, providing us with the use of theoretical tools we would not otherwise have at our disposal.  With high-energy nuclear beams and the straightforward access to kinematic variables offered by DIS, the EIC would be an ideal facility at which to study the saturation regime in QCD.  For a more complete discussion of the physics possible with $e$+A collisions at the EIC, see \cite{eAPaper}.

\section{Precision Studies of Nucleon Spin Structure}

Similar to the case of nuclear DIS experiments, polarized DIS experiments have thus far utilized fixed targets and therefore explored only a restricted kinematic region.  In particular, knowledge of the gluon helicity distribution from polarized DIS experiments is limited by the narrow range in $Q^2$ of the present data.  The EIC would be able to access the polarized gluon distribution down to $x$ values of several times $10^{-4}$ through various inclusive and semi-inclusive measurements, with uncertainties small enough to gain sensitivity to the functional form for the first time.

In recent years an increasing amount of attention has been focused on the transverse spin structure of the proton.  Large transverse single-spin asymmetries have been observed over a broad range of center-of-mass energies, which are believed to be due to spin-orbit effects in the nucleon itself and/or in the process of fragmentation into hadrons.  In order to study the orbital angular momentum of the partons within a hadron, one clearly needs to go beyond the framework of collinearly factorized pQCD.  A wealth of high-energy data relevant to transverse-momentum-dependent, i.e.\ non-collinear, distribution and fragmentation functions has become available in the last few years, and simultaneous fits to data from a variety of experiments are beginning to be performed \cite{CollinsFit,SiversFit2,SiversFit}.  The high-energy DIS data sensitive to transverse-momentum-dependent distributions that the EIC could provide would help bring this field to maturity.

The ultimate way to map out the spin-orbit structure of the nucleon may lie in the measurement of generalized parton distributions (GPD's).  GPD's simultaneously describe the position and momentum distributions of partons within the nucleon, representing the closest analog to the phase-space density allowed by the uncertainty principle.  Further detail on the opportunities to study GPD's at the EIC can be found in \cite{GPD}.

\section{Collider Design Concepts}\label{collider}

\begin{figure}[bt]
     \begin{center}
    \includegraphics[width=0.5\textwidth]{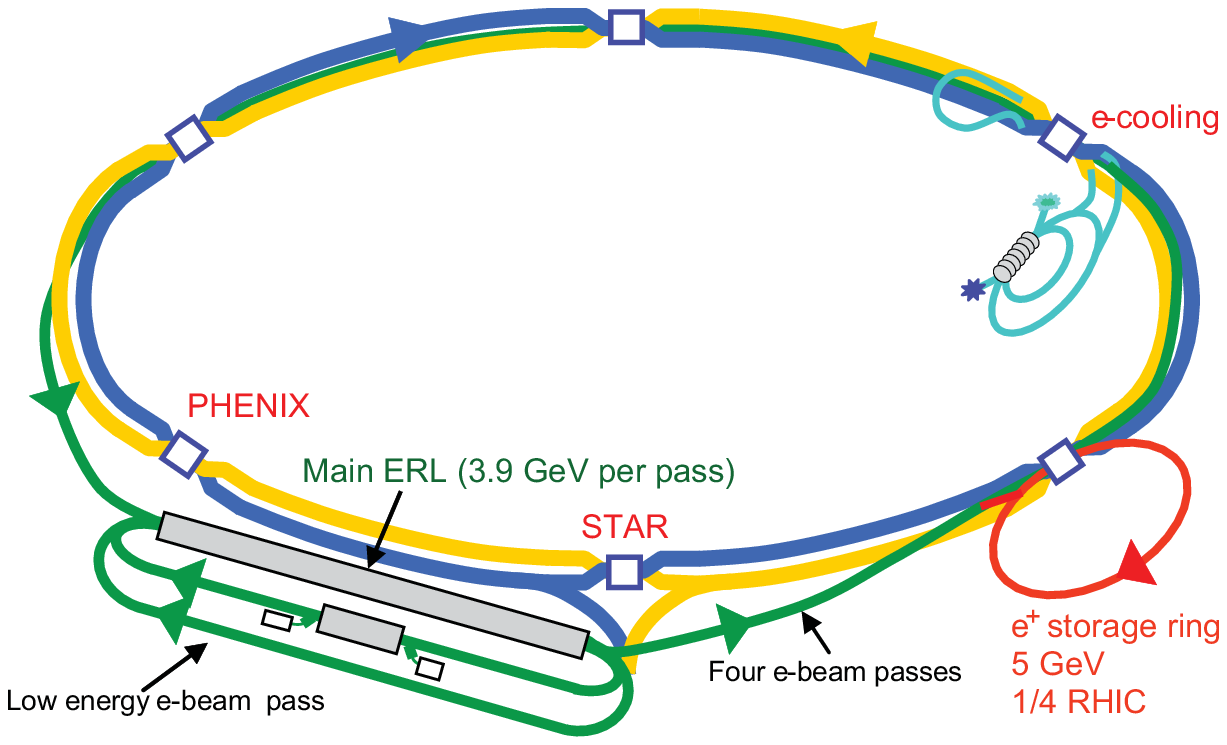}
    \hspace{0.05\textwidth}
    \includegraphics[width=0.38\textwidth]{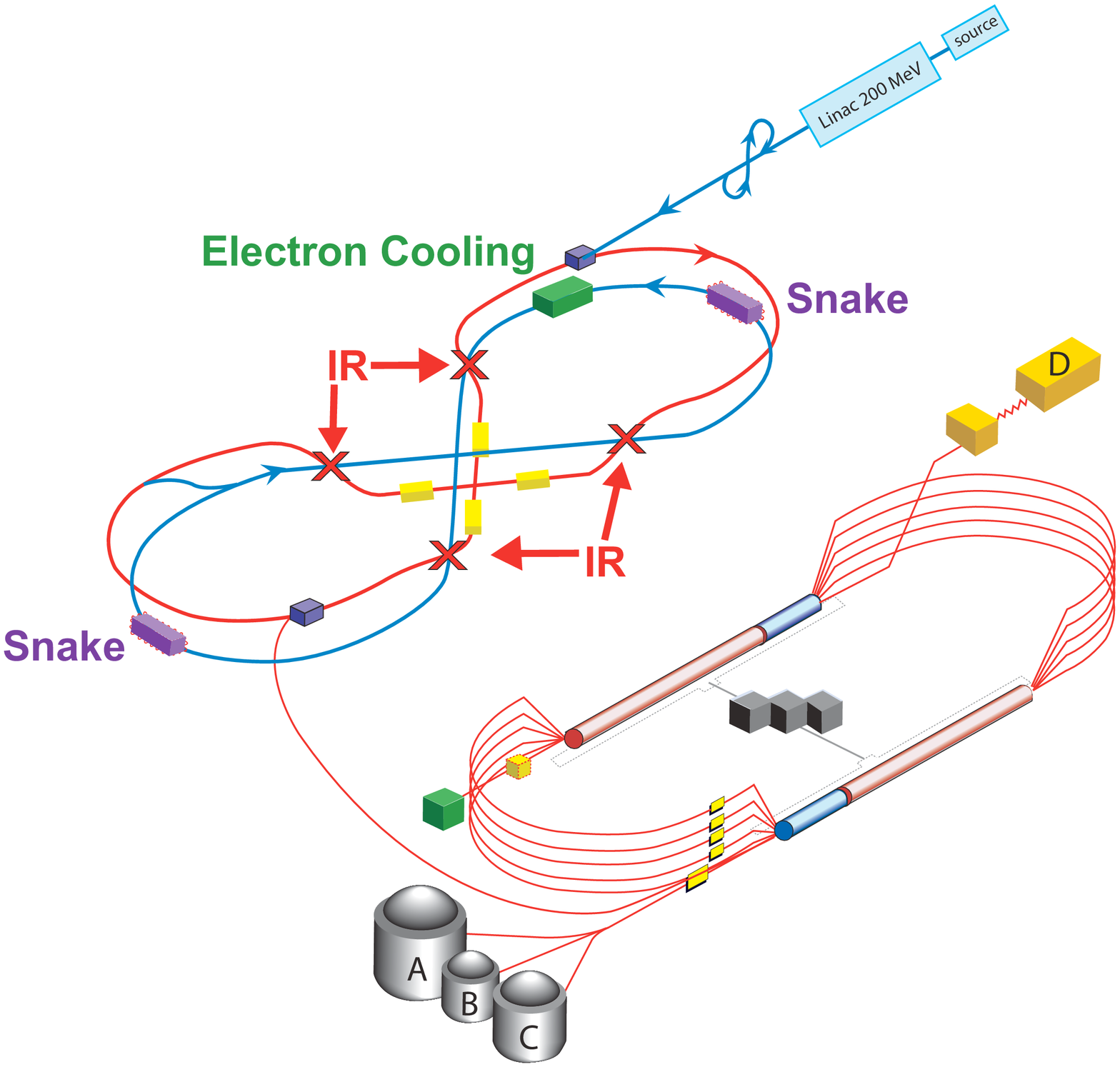}
    \vspace{-3mm}
     \end{center}
     \caption{Schematic designs of the eRHIC (left) collider at BNL
         based on an energy recovery linac and ELIC (right) at Jefferson Lab.}
    \label{fig:eic_concepts}
\end{figure}

There are two EIC design concepts currently under study.  One proposal is to add an electron facility to the existing RHIC collider at BNL (eRHIC); the other is to add a hadron facility to the upgraded CEBAF accelerator at Jefferson Lab (ELIC).  The schematic layout of both designs can be seen in Fig.~\ref{fig:eic_concepts}.

Basic machine requirements include variable center-of-mass energy from 20-100 GeV, high luminosity ($>10^{32}$ cm$^{-2}$s$^{-1}$), the ability to accelerate ions up to A=208, and the ability to maintain polarized electron, proton, and light ion beams.  Greater detail regarding the accelerator design proposals can be found in \cite{eRHIC,ELIC}.

\section{Summary}\label{summary}

The proposed Electron-Ion Collider promises to be a formidable machine for the study of QCD matter.  As a unique facility performing deep-inelastic scattering at high energies of electrons off of heavy nuclear as well as polarized proton and light ion beams, it will offer a wide range of physics opportunities, from exploring the nature of strong color fields in nuclei to precisely imaging the sea quarks and gluons within the nucleon to map out their spin, flavor, and spatial distributions.

\section*{Note}
\begin{notes}
\item[a]
E-mail: caidala@bnl.gov
\end{notes}

\vfill\eject
\end{document}